\global\def\draftcontrol{0}  
  
   \def\versionno{momentum in 0A}  
  
\catcode`\@=11  
  
\expandafter\ifx\csname draftcontrol\endcsname\relax\global\def\draftcontrol{0}  
\fi  
  
{\count255=\time\divide\count255 by 60  
\xdef\hourmin{\number\count255}  
\multiply\count255 by-60\advance\count255 by\time  
\xdef\hourmin{\hourmin:\ifnum\count255<10 0\fi\the\count255}}  
\def\draftdate{\number\month/\number\day/\number\year\ \ \ \hourmin }  
  
\newcommand\makepapertitle{\par  
  \begingroup  
    \renewcommand\thefootnote{\@fnsymbol\c@footnote}%
    \def\@makefnmark{\rlap{\@textsuperscript{\normalfont\@thefnmark}}}%
    \long\def\@makefntext##1{\parindent 1em\noindent  
            \hb@xt@1.8em{%
                \hss\@textsuperscript{\normalfont\@thefnmark}}##1}%
     \newpage  
     \global\@topnum\z@   
     \@makepapertitle  
     \thispagestyle{empty}\@thanks  
  \endgroup  
  \setcounter{footnote}{0}%
  \global\let\thanks\relax  
  \global\let\makepapertitle\relax  
  \global\let\@makepapertitle\relax  
  \global\let\@thanks\@empty  
  \global\let\@author\@empty  
  \global\let\@date\@empty  
  \global\let\@title\@empty  
  \global\let\title\relax  
  \global\let\author\relax  
  \global\let\date\relax  
  \global\let\and\relax  
  \def\version{\let\version\@version\@gobble}  
}  
\def\@makepapertitle{%
  \newpage  
   \ifnum\draftcontrol=1 {}  
   \version\versionno  
   \vskip 3em%
   \else  
   \hfill\hbox to 3cm {\parbox{4cm}{\@pubnum}\hss}%
   \vskip 3em%
   \fi  
   \begin{center}%
   \let \footnote \thanks  
     {\LARGE {\@title}}%
     \vskip 1.5em%
     {\normalsize
       \lineskip .5em%
       \begin{tabular}[t]{c}%
         \@author  
       \end{tabular}\par}%
     \vskip 1.5em%
     {\@bstract}%
     \end{center}%
     \vskip 1.5em  
     \@date%
   \par  
}  
  
\gdef\@pubnum{}  
\def\pubnum#1{%
  \gdef\@pubnum{#1}}  
  
\gdef\@bstract{}  
\def\Abstract#1{%
  \gdef\@bstract{%
   \parbox{\textwidth-0pc}{%
   \centerline{\bf Abstract}\penalty1000%
\noindent
\renewcommand\baselinestretch{1.0}%
{#1}}}  
}  
  
\def\ps@paper{\let\@mkboth\@gobbletwo%
     \ifnum\draftcontrol=1  
        \def\@oddfoot{\hbox to \textwidth{\tiny \versionno \hfil\tiny\draftdate}%
        \hskip -\textwidth \hbox to \textwidth{\hfil\rm\thepage\hfil}}%
     \else\def\@oddfoot{\hbox to \textwidth{\hfil\rm\thepage\hfil}}  
     \fi  
     \let\@evenfoot\@oddfoot  
}  
  
\def\@version#1{\ifnum\draftcontrol=1  
\typeout{}\typeout{#1}\typeout{}  
\vskip3mm\centerline{\hbox{\fbox{\normalsize{\tt DRAFT -- #1 -- }  
                   {\draftdate}}}}\vskip3mm  
\fi}  
\let\version\@version  
\long\def\eqlabel#1{\ifnum\draftcontrol=1  
                    \tag@false  
                    \tag*{(\theequation) \hbox to -0.2cm{\hspace{0cm}\small{#1}\hss}}  
                    \refstepcounter{equation}  
                    \edef\@currentlabel{\theequation}  
                    \ltx@label{#1}          
                    \else  
                    \label{#1}  
                    \fi  
                    }  
\let\st@bibitem\@bibitem  
\let\st@lbibitem\@lbibitem  
\ifnum\draftcontrol=1  
  \def\@bibitem#1{%
    \st@bibitem{#1}\a@@label{#1}\ignorespaces}  
  \def\@lbibitem[#1]#2{%
    \st@lbibitem[#1]{#2}\a@@label{#2}\ignorespaces}  
  \def\a@@label#1{%
    \gdef\a@lab{\smash{\normalfont\small#1}}  
    \ifvmode  
      \if@inlabel  
        \global\setbox\@labels\hbox{%
          \llap{\a@lab\let\a@lab\relax  
                \kern\@totalleftmargin\kern\marginparsep}%
          \box\@labels}%
      \fi  
    \fi}  
\fi  
  
\documentclass[12pt,letterpaper]{article}  
  
\usepackage{amsmath,amssymb,array,calc,rotating,epsfig,psfrag}  
\usepackage[nosort]{cite}  
  
\ifnum\draftcontrol=1  
\fi  
  
\renewcommand\baselinestretch{1.25}  
\renewcommand\section{\@startsection {section}{1}{\z@}%
                                   {-3.5ex \@plus -1ex \@minus -.2ex}%
                                   {2.3ex \@plus.2ex}%
                                   {\normalfont\large\bfseries}}  
\renewcommand\subsection{\@startsection{subsection}{2}{\z@}%
                                   {-3.25ex\@plus -1ex \@minus -.2ex}%
                                   {1.5ex \@plus .2ex}%
                                   {\normalfont\normalsize\bfseries}}  
\renewcommand\subsubsection{\@startsection{subsubsection}{3}{\z@}%
                                   {-3.25ex\@plus -1ex \@minus -.2ex}%
                                   {1.5ex \@plus .2ex}%
                                   {\normalfont\normalsize\it}}  
\renewcommand\paragraph{\@startsection{paragraph}{4}{\z@}%
                                   {-3.25ex\@plus -1ex \@minus -.2ex}%
                                   {1.5ex \@plus .2ex}%
                                   {\normalfont\normalsize\bf}}  
  
\def\revise#1       {\raisebox{-0em}{\rule{3pt}{1em}}%
                     \marginpar{\raisebox{.5em}{\vrule width3pt\  
                     \vrule width0pt height 0pt depth0.5em  
                     \hbox to 0cm{\hspace{0cm}{%
                     \parbox[t]{4em}{\raggedright\footnotesize{#1}}}\hss}}}}

\def\calc         {{\cal C}}

\def\calj         {{\cal J}}

\def\calo         {{\cal O}}

\def\del          {\partial}

\def\tr           {\mathop{\rm Tr}}

\def\half{{\frac12}}

\def\sqr#1#2{{\vcenter{\vbox{\hrule height.#2pt  
 \hbox{\vrule width.#2pt height#1pt \kern#1pt  
 \vrule width.#2pt}\hrule height.#2pt}}}}

\def\a{\alpha}  
\def\b{\beta}  
\def\r{\rho}

\def\m{\mu}  
\def\g{\gamma}  
\def\l{\lambda}  
\def\n{\nu}  
\def\bn{\bar{\nu}}  
\def\bm{\bar{\mu}}  
  
\catcode`\@=12

\usepackage{color}

\begin{document}  
  
  
  

\newcommand{\be}{\begin{equation}}  
\newcommand{\ee}{\end{equation}}  
\newcommand{\beq}{\begin{equation}}  
\newcommand{\eeq}{\end{equation}}  
\newcommand{\ba}{\begin{eqnarray}}  
\newcommand{\ea}{\end{eqnarray}}  
\newcommand{\nn}{\nonumber}  
  
\def\vol{\bf vol}  
\def\Vol{\bf Vol}  
\def\del{{\partial}}  
\def\vev#1{\left\langle #1 \right\rangle}  
\def\cn{{\cal N}}  
\def\co{{\cal O}}  
\def\IC{{\mathbb C}}  
\def\IR{{\mathbb R}}  
\def\IZ{{\mathbb Z}}  
\def\RP{{\bf RP}}  
\def\CP{{\bf CP}}  
\def\Poincare{{Poincar\'e }}  
\def\tr{{\rm tr}}  
\def\tp{{\tilde \Phi}}  
\def\Y{{\bf Y}}  
\def\te{\theta}  
\def\bX{\bf{X}}  
  
\def\TL{\hfil$\displaystyle{##}$}  
\def\TR{$\displaystyle{{}##}$\hfil}  
\def\TC{\hfil$\displaystyle{##}$\hfil}  
\def\TT{\hbox{##}}  
\def\HLINE{\noalign{\vskip1\jot}\hline\noalign{\vskip1\jot}} 
\def\seqalign#1#2{\vcenter{\openup1\jot  
  \halign{\strut #1\cr #2 \cr}}}  
\def\lbldef#1#2{\expandafter\gdef\csname #1\endcsname {#2}}  
\def\eqn#1#2{\lbldef{#1}{(\ref{#1})}%
\begin{equation} #2 \label{#1} \end{equation}}  
\def\eqalign#1{\vcenter{\openup1\jot   }}  
\def\eno#1{(\ref{#1})}  
\def\href#1#2{#2}  
\def\half{{1 \over 2}}

\def\ads{{\it AdS}}  
\def\adsp{{\it AdS}$_{p+2}$}  
\def\cft{{\it CFT}}  
  
\newcommand{\ber}{\begin{eqnarray}}  
\newcommand{\eer}{\end{eqnarray}}  
  
\newcommand{\bea}{\begin{eqnarray}}  
\newcommand{\eea}{\end{eqnarray}}  
  
\newcommand{\beqar}{\begin{eqnarray}}  
\newcommand{\cN}{{\cal N}}  
\newcommand{\cO}{{\cal O}}  
\newcommand{\cA}{{\cal A}}  
\newcommand{\cT}{{\cal T}}  
\newcommand{\cF}{{\cal F}}  
\newcommand{\cC}{{\cal C}}  
\newcommand{\cR}{{\cal R}}  
\newcommand{\cW}{{\cal W}}  
\newcommand{\eeqar}{\end{eqnarray}}  
\newcommand{\lm}{\lambda}\newcommand{\Lm}{\Lambda}  
\newcommand{\eps}{\epsilon}  
  
  
\newcommand{\nonu}{\nonumber}  
\newcommand{\oh}{\displaystyle{\frac{1}{2}}}  
\newcommand{\dsl}  
  {\kern.06em\hbox{\raise.15ex\hbox{$/$}\kern-.56em\hbox{$\partial$}}}  
\newcommand{\as}{\not\!\! A}  
\newcommand{\ps}{\not\! p}  
\newcommand{\ks}{\not\! k}  
\newcommand{\D}{{\cal{D}}}  
\newcommand{\dv}{d^2x}  
\newcommand{\Z}{{\cal Z}}  
\newcommand{\N}{{\cal N}}  
\newcommand{\Dsl}{\not\!\! D}  
\newcommand{\Bsl}{\not\!\! B}  
\newcommand{\Psl}{\not\!\! P}  
\newcommand{\eeqarr}{\end{eqnarray}}  
\newcommand{\ZZ}{{\rm \kern 0.275em Z \kern -0.92em Z}\;}  
  
\def\s{\sigma}  
\def\a{\alpha}  
\def\b{\beta}  
\def\r{\rho}  
\def\d{\delta}  
\def\g{\gamma}  
\def\G{\Gamma}  
\def\ep{\epsilon}  
\makeatletter \@addtoreset{equation}{section} \makeatother  
\renewcommand{\theequation}{\thesection.\arabic{equation}}  

\def\be{\begin{equation}}  
\def\ee{\end{equation}}  
\def\bea{\begin{eqnarray}}  
\def\eea{\end{eqnarray}}  
\def\m{\mu}  
\def\n{\nu}  
\def\g{\gamma}  
\def\p{\phi}  
\def\L{\Lambda}  
\def \W{{\cal W}}  
\def\bn{\bar{\nu}}  
\def\bm{\bar{\mu}}  
\def\bw{\bar{w}}  
\def\ba{\bar{\alpha}}  
\def\bb{\bar{\beta}}

\begin{titlepage}  
  
\version\versionno  
  
\leftline{\tt hep-th/0507061}  
  
\vskip -.8cm  
  
\rightline{\small{\tt MCTP-05-84}}

\vskip 1.7 cm  
  
\centerline{\bf \large  Condensing Momentum Modes in 2-d 0A String Theory   
with Flux}

\vskip .2cm \vskip 1cm {\large } \vskip 1cm  
  
\centerline{\large  Leopoldo A. Pando Zayas and  Diana Vaman  }  
  
\vskip .5cm  
\centerline{\it Michigan Center for Theoretical  
Physics}  
\centerline{ \it Randall Laboratory of Physics, The University of  
Michigan}  
\centerline{\it Ann Arbor, MI 48109-1040}  
  
\vspace{1cm}  
  
\begin{abstract}  
We use a combination of conformal perturbation theory techniques   
and matrix model results to study the effects of perturbing by momentum modes   
two dimensional type 0A strings with non-vanishing Ramond-Ramond   
(RR) flux. In the limit of large RR flux (equivalently, $\mu=0$)   
we find an explicit analytic form of the genus zero partition function  
in terms of the RR flux $q$ and the momentum modes coupling constant $\alpha$.   
The analyticity of the partition function  
enables us to go beyond the perturbative regime and, for   
$\alpha\gg q$, obtain the   
partition function in a background corresponding to the momentum modes   
condensation.  
For momenta such that $0<p<2$ we find no obstruction to   
condensing the momentum modes in the phase diagram of the partition function.  
  
\end{abstract}

  

\end{titlepage}  
  
  
  
\section{Introduction }  
The open/closed string correspondence is one of the fundamental  
concepts in the modern understanding of string theory. This  
correspondence provides, in various cases, a non-perturbative  
definition of string theory.   
  
The AdS/CFT  
correspondence is perhaps one of the best studied instances of the  
open/closed string correspondence. Another very important case is  
string theory in two dimensions where the open string side of the  
correspondence  is described via a matrix model. The main attraction  
of the open/closed string correspondence in two dimensions resides in  
the ability to obtain exact results on both sides of the  
correspondence.   
  
The simplest case of two-dimensional duality is provided by the $c=1$  
model. The open string side is described by an exactly solvable random  
matrix model with inverted harmonic potential. The closed string side  
is a Liouville theory which has  
been solved using the conformal bootstrap.   
  
Recently, two new non-supersymmetric two-dimensional string theories   
have been formulated and their corresponding matrix models identified \cite{0b,chat}. In   
the spirit of the renormalization group flow, it is natural to study   
the deformation of the above correspondence, that is to study the   
relationship between the two descriptions after adding operators or   
expectation values to these theories.   
  
In this paper we study the deformation of two-dimensional type 0A  
string theory by momentum modes. We employ a technique successfully  
applied to the $c=1$ model by G. Moore in \cite{mooresg} (see also   
\cite{kutasovsg}). This technique uses a combination of conformal   
perturbation theory and matrix model results. In recent years the   
beautiful results of \cite{mooresg,kutasovsg} have been reproduced   
and improved using alternative techniques \cite{bk,kkk}. In particular, the study   
of adding momentum and winding perturbations to the $c=1$ model has   
explicitly revealed the rich mathematical structure of integrable   
systems in these models \cite{kostov,akk,kostov2,time}. Various physical aspects   
related to phase transitions have been confirmed and reinterpreted.   
Most remarkably among them are the connection with the Euclidean two-dimensional   
black hole \cite{kkk} and to time-dependent backgrounds   
\cite{time,time1,time2,time3}.   
  
The idea that some two-dimensional black holes admit a matrix model  
description has a long history. A prominent role has been played by a  
deformation of the inverted harmonic oscillator matrix model due to  
Jevicki and Yoneya (JY) \cite{jy}. This precise matrix model has resurfaced  
recently as it describes type 0A string theory in  
the presence of D-branes \cite{chat}.   
  
Indeed, there was some evidence that certain aspects of the deformed  
matrix model match their counterpart in the 0A two-dimensional black hole  
\cite{gtt}. A closer examination, however, showed that the  
thermodynamics of two-dimensional type 0A black holes does not match that of the  
deformed matrix model \cite{jld,danielsson,jb}. It was suggested in   
\cite{jld}, that the two-dimensional 0A black hole  has properties similar to   
that of a different deformation of the c=1 matrix model considered by   
Boulatov and Kazakov \cite{bk} and applied to the $c=1$ black hole   
in \cite{kkk}. Kazakov and Tseytlin \cite{kt} compared the   
matrix model deformed by vortices with the exact two-dimensional black hole   
obtained in \cite{dvv} and found some qualitative agreement.   
Despite much effort, the existence of a   
direct correspondence between two-dimensional Lorentzian black holes   
and matrix models is still  under scrutiny \cite{blackhole}.   
  
Irrespectively of the ultimate relationship of perturbed two-dimensional string theories with two-dimensional black holes, our   
work is interesting in its own right as it provides an explicit expression for the partition function of   
the Jevicki-Yoneya (JY) model in the presence of momentum modes.    
{}From the matrix model point of  
view we are computing the effect of adding momentum modes  in a model that  
provides a non-perturbatively calculable unitary S matrix \cite{igor}. Other interesting nonperturbative aspects  
have been discussed in, for example, \cite{smatrix,kapustin,johanes,kms}.  
  
The paper is organized as follows.   
In section \ref{review} we review the work of Moore in   
\cite{mooresg}, outlining the strategy that we will follow   
and introducing most of the notation. Section \ref{0A}   
contains our main result, the   
partition function of the two-dimensional   
type 0A string theory perturbed by momentum modes.   
Using the dual matrix model description in terms of free fermions,   
the deformed JY model, we find an   
explicit analytical expression for the genus zero partition function,   
in the limit of a vanishing Fermi energy.    
In Section \ref{phase}  we analyze the phase diagram in terms  
of the three parameters: the momentum $p$, the RR flux $q$ and the   
coupling constant of the momentum modes $\alpha$. We   
conclude in Section \ref{conclusions} with comments on the   
approximations used in this paper and some open problems. In appendix \ref{app} we apply the Lagrange   
Inversion Formula to obtain and analytic expression for the partition function and comment on its analytic continuation.   
  
\section{Review of the gravitational Sine-Gordon model ($c=1$ perturbed by momentum modes)}\label{review}  
In this section we review Moore's analysis \cite{mooresg}. Similar calculations were also   
performed in \cite{kutasovsg}, and in \cite{kkk}. This review should  
provide much of the notation and the logical framework we will use in  
the next section.   
  
The Sine-Gordon (SG)  model coupled to two-dimensional gravity is given by the following action:   
\begin{eqnarray}  
\label{sggravity}  
S(\mu,\lambda)&=& \int d^2 z \sqrt{\hat{g}} \left( \frac{1}{8\pi} (\nabla \phi)^2 +  
\frac{\mu}{8\pi \g^2}e^{\g \phi} + \frac{Q}{8\pi}\phi  
R({\hat{g}})\right)\nonumber \\  
&+& \int d^2 z \sqrt{\hat{g}}\left(\frac{1}{8\pi} (\nabla X)^2 + \lambda e^{\xi \phi} \cos  
(p X)\right),  
\end{eqnarray}  
where $\hat{g}$ is a background metric, $R$ its curvature, $\mu$ the cosmological constant.  
  
There are various motivations for considering the above problem.   
Coupling the SG model with two-dimensional gravity helps understand aspects of   
the SG model as a quantum field theory. For example, properties like   
the existence of certain  RG trajectories are expected to be insensitive to coupling to gravity.   
  
The interpretation that is more along the lines of our interest here is as follows.  
Consider the free action of two uncompactified  
real fields $\phi$ and $X$:  
\begin{eqnarray}  
S_{Liouville}+S_{Gaussian}&=& \int d^2 z \sqrt{g}\left(\frac{1}{8\pi}  
(\nabla \phi)^2 +\frac{Q}{8\pi} \phi R(g) \right) \nonumber\\  
&+& \int d^2 z \sqrt{g} \frac{1}{8\pi} (\nabla X)^2.  
\end{eqnarray}  
It is natural to consider, in the spirit of RG, perturbing this   
action by an operator of the form  
\be   
\sum\limits_i e^{\xi_i \phi}   
\calo_i,   
\ee   
where $\calo_i$ are operators of the $c=1$ Gaussian   
model, the values of $\xi_i$ are selected such that the dressed operator  
has conformal dimension one. For the choice of $\calo_i$ made in \ref{sggravity}, the   
problem formulated above is that of {\it perturbing the $c=1$ model   
by momentum modes.}   
  
The central object is the partition function defined as \be   
Z=\vev{e^{-S(\mu,\lambda)}}. \ee In the limit were $\lambda$ is very   
small the techniques of conformal perturbation theory become   
available to us. Namely, we can think of $Z$ as a series expansion   
of the form :   
\be   
Z=\sum\limits_{n=0}^\infty \frac{\lambda^n}{n!}\langle   
\left(\cos p X  e^{\xi \phi}\right)^n\rangle_{\lambda=0} =   
\sum\limits_{n=0}^\infty\frac{(\frac12\l^2)^n}{(n!)^2}\langle   
(e^{ipX+\xi \phi})^n( e^{-ipX+\xi \phi})^n\rangle_{\l=0}.   
\ee   
As can be seen from the above expression, we will be interested in   
correlators of the form  
\begin{equation}  
\eqlabel{correlator}  
\vev{\prod V_{q_i}e^{\frac12 \l (V_p +V_{-p})}} \equiv \sum\limits_{n_1,  
n_2 \ge 0}\frac{\l^{n_1+n_2}}{2^{n_1+n_2}n_1! n_2!} \vev{ \prod V_{q_i}  
(V_p)^{n_1} (V_{-p})^{n_2}}.  
\ee  
where   
\be   
V_p= \int d^2 z\sqrt{\hat{g}}e^{\xi \phi} e^{ipX}.   
\ee  
A remarkable aspect of the duality between two-dimensional string theories and matrix models is that one can actually compute   
all the correlators in \ref{correlator}  using matrix models \cite{mpr,mooresg}.   
  
In the matrix model framework it is convenient to work with rescaled  
operators and coupling:  
\be   
\calj_{\pm p} =\frac{\Gamma(p)}{\Gamma(-p)}V_{\pm p} =   
\frac{\Gamma(p)}{\Gamma(-p)}\int d^2 z \sqrt{g} e^{\xi \phi} e^{\pm   
ip X}, \quad \a= \frac{\Gamma(-p)}{\Gamma(p)} \l.   
\ee  
Note that with the above notation the partition function takes the  
simple form of: \be  
 Z =  \vev{e^{\a\calj_p + \a\calj_{-p}}}.  
\ee  
In principle, one could proceed to evaluate  
$\vev{\calj_p^n\calj_{-p}^n}$ using the prescription of \cite{mpr}.   
However,  a simpler way to evaluate it is by inserting a zero   
momentum operator inside the correlators. The simplification   
comes  about due to a couple of observations made in \cite{mpr}.   
First, note that introducing $\calj_0$ into a correlator is   
equivalent  to differentiating the correlator  with respect to   
$\mu$. Second, since $\mu$ always enters as $p+i\mu$ we see that   
differentiation with respect to $\mu$ is equivalent  to   
differentiation with respect to $p$. Inspecting the general form   
of the amplitudes in \cite{mpr} we see that differentiating with   
respect to momentum turns the $\theta$-functions  into   
$\delta$-functions making integration a simple task. In other    
words, inserting $\calj_0$ into the amplitudes has the advantage of   
turning a complicated integration into a manageable combinatorial   
formula presented in appendix A of \cite{mooresg}:  
\begin{eqnarray}  
\label{an}  
A_n(\mu, p) &\equiv & \mu^{-np} \vev{\calj_0\calj_p^n \calj_{-p}^n} \\  
&=&i (-1)^n (n!)^2 \sum\limits_{k=1}^n \frac{(-1)^k}{k} \sum\limits_{a_i,  
b_i} \left(\prod\limits_{i=1}^k b_i^2 - \prod\limits_{i=1}^k  
a_i^2\right) \calc(a_1, \ldots, b_k) \prod\limits_{i=1}^k  
\frac{R_{a_ip}R^{*}_{b_ip}}{(a_i)!{}^2 (b_i)!{}^2 },\nonumber  
\end{eqnarray}  
where $R_p$ is the bounce factor of $c=1$ for momentum $p$. The sum is over all partitions of $n=a_1+b_1+\ldots + a_k + b_k$ with  
$a_i, b_i \ge 0$ such that  
\be  
\calc(a_1, \ldots, b_k) \equiv  
\frac{1}{(a_1+b_1)(b_1+a_2)(a_2+b_2)\ldots(a_k+b_k)(b_k + a_1)},  
\ee  
is nonzero. The last step in recovering the amplitudes that enter in the partition  
function involves integrating over $\mu$.   
  
Since we are interested in the genus zero partition function, it is   
worth considering the asymptotic form of the bounce factor for the   
$c=1$ model   
\begin{equation}  
\eqlabel{bounceperturbative}  
R_p^{\mu\to \infty}=  \exp\bigg[ p\psi +  
\sum\limits_{n\ge 1} \frac{i^n  
p^{n+1}}{(n+1)!}\left(\frac{d}{d\mu}\right)^n (\log\mu + \psi)\bigg]  
= 1+ \frac{ip^2}{\mu} +\ldots,  
\end{equation}  
where,  
\be \label{psi}  
\psi\equiv \sum\limits_{k\ge   
1}\frac{(-1)^kB_{2k}}{2k}(1-2^{-2k+1})\frac{1}{\mu^{2k}},  
\ee  
and $B_{2k}$ are the Bernoulli numbers. With this expression for   
the bounce factor Moore finds that the genus zero amplitude is given   
by  
\be \label{c1cor}  
A_n^{h=0}(\mu,p)\equiv \mu^{-np} \langle {\cal J}_0{\cal   
J}_p^n{\cal J}_{-p}^n \rangle   
=n!\mu^{-2n+1}\frac{\Gamma(n(1-p)+n-1)}{\Gamma(n(1-p)+1)}(1-p)^n   
p^{2n}\, .   
\ee   
Integrating with respect to $\mu$ we find that the needed   
correlators are:  
\be\label{tntn}  
\vev{\calj_p^n\calj_{-p}^n}= -\mu^{np -2n +2}n! p^{2n}(1-p)^n\frac{\G(n(1-p)+n-2)}{\G(n(1-p)+1)}.  
\ee  
An insightful way of assembling the answer was presented in   
\cite{kkk}\footnote{In appendix \ref{app} we included a derivation of this formula as an   
application of the Lagrange Inversion Formula.}. Using that   
\be\label{sumformula}  
\sum_{n\geq 1}\frac{\Gamma(na+b-1)}{n!\Gamma(n(a-1)+b)}(-z)^n=  
\frac{(1-s)^{b-1}}{b-1}, \quad {\rm where} \frac{s}{(1-s)^a}\equiv z,   
\ee  
one obtains an expression for the susceptibility $\chi=\partial^2_\mu  
Z$ of the form:  
\be  
\chi=-\ln \mu + \ln (1-s),  
\ee  
where in this case   
\be  
z=\alpha^2\, p^2(p-1)\m^{p-2}, \qquad a =  2-p, \qquad b=1.  
\ee  
This expression for the susceptibility can be rewritten as   
\be  
\mu\,  e^{\chi}+\alpha^2 \, p^2\, (p-1) e^{(2-p)\chi}=1.  
\ee  
The main advantage of the above expression is that it allows finding the   
large $\a$ behavior in the region where $\mu$ can be turned  
off. As a bonus, the KPZ scaling in the new   
coupling \cite{mooresg,kkk} can be verified automatically.   
Namely, we find that in this limit  
\be  
\eqlabel{mu0}  
\chi_{\mu=0}=-\frac{1}{2-p}\ln \alpha^2 p^2 (p-1).  
\ee  
  
\section{Type 0A perturbed by momentum modes}\label{0A}  
In this section we discuss perturbing two-dimensional type 0A string  
theory by momentum modes following the techniques of  
\cite{mooresg,kutasovsg}.   
  
The matrix model description of type 0A with $q$ unit of   
fluxes\footnote{ Recently, a clarification of the meaning of $q$   
has been given \cite{ms} as the sum of the two distinct fluxes and was  
denoted by $\hat{q}$. A similar interpretation was put forward  
previously in \cite{jb} based on a thermodynamical analysis of the low  
energy supergravity  action.}  was  
recently established  \cite{chat} to be the Jevicki-Yoneya (JY)   
matrix model \cite{jy}. Essentially, this is a matrix model with the   
following potential:   
\be\label{defmatr}  
V(x)=-\frac{x^2}{2}+  
\frac{q^2-1/4}{2x^2}.  
\ee  
This model was solved a decade ago, its non-perturbative S-matrix and  
the explicit form for   
some of the amplitudes were discussed in \cite{igor,dr}.   
  
Our goal is to compute the partition function of the two-dimensional type 0A string   
theory, {\it with non-vanishing RR flux}, and in the presence of   
momentum perturbations. Formally, we would like to compute:   
\be  
\eqlabel{z0a} Z=\langle \exp(\lambda \cos(p X)e^{\xi\phi})\rangle_{0A}\, . \ee  
In practice, however, we lack a worldsheet action analogous to \ref{sggravity}.   
Nevertheless, via the string/matrix model correspondence we take the   
momentum correlators in two-dimensional type 0A to be those of momentum operators in the JY matrix model.   
Then, we interpret \ref{z0a} as defined with the momentum correlators computed using the matrix models.   
Within conformal perturbation theory,   
the partition function is composed of building blocks similar to the $c=1$ case, that is,    
the partition function is obtained from correlators of the form  
$\langle {\calj}_p^n{\calj}_{-p}^n\rangle$.  
  
\subsection{Bounce factor}  
  
Our starting point is the bounce factor of the  JY matrix model:   
\be \eqlabel{0ar}   
R(p)=\bigg(\frac{4}{q^2+\mu^2-1/4}\bigg)^{p/2}\frac{\Gamma(\frac12(1+q+p-i\mu))}   
{\Gamma(\frac12(1+q-p+i\mu))}.   
\ee  
As in the $c=1$ model, we build the genus zero partition function in the   
presence of  momentum modes perturbatively in the coupling constant $\lambda$,   
with $\lambda\ll \mu$. This amounts to expanding the bounce factor  
as a series in inverse powers of $\mu$.   
However, as a result of having $\mu\to\infty$ in this expansion,   
the dependence on the RR flux will   
be washed out. To avoid this we choose an alternative limit in which the   
RR flux scales with $\mu$:  
\be  
q=\mu f\, .  
\ee  
Introducing the notation  
\be \label{mu1mu2}  
\mu_1=\mu(if + a), \qquad \mu_2 =\mu (if   
-a)\, ,   
\ee  
 up to an overall $p$-independent   
phase we can rewrite the bounce factor (\ref{0ar}) as   
\bea \label{bounce0a}   
R(p)&=&(\mu_1\mu_2)^{p/2}\bigg(\frac{16}{16\mu_1\mu_2-1}\bigg)^{p/2}   
\exp\bigg(p/2\psi(\mu_1)+p/2\psi(\mu_2) \nonumber\\&+&\sum_{n\geq   
1}\frac{p^{n+1}}   
{2^{n+1}(n+1)!}(\partial_{\mu_1}^n\ln\mu_1+(-)^n\partial_{\mu_2}^n\ln\mu_2   
+\partial_{\mu_1}^n\psi(\mu_1)+(-)^n\partial_{\mu_2}^n\psi(\mu_2))\bigg)\, ,  
\nonumber\\  
\eea  
where $\psi$ is defined as in (\ref{psi}).   
In (\ref{mu1mu2}), $a$ is a marker introduced for later   
purposes, with $a=1$ its canonical value. Introducing $a$ allows, among other things, to turn off   
$\mu$ (by setting $a=0$) without turning off the RR flux at the same time. Also, the limit   
$f=0$ is expected to bring us back to the $c=1$ bosonic string.  
  
Expanding in the first few orders in inverse powers of   
$\mu$ we get:   
\bea   
\label{bounce0aperturbative}  
R(p)&=& 1+\frac{i}{2}\frac{a}{\mu(f^2+a^2)}p^2 \nonumber \\  
&-& \frac{p}{24\mu^2 (f^2+a^2)^2}\left(-7f^2 +a^2 +4p^2f^2 -4p^2a^2+3p^3a^2\right) \nonumber  \\  
&-&\frac{i}{48}\frac{a}{\mu^3(f^2+a^2)^3} p^2 \left(   
-8a^2+pa^2+4p^2 a^2-4p^3 a^2+p^4 a^2  
\right. \nonumber \\  
&+&\left. 24 f^2-7pf^2-12p^2 f^2+4p^3 f^2\right) +  
\calo(\mu^{-4})\, .  
\eea  
Some consistency checks are in order. First, note that for $f=0$ we   
basically return to the $c=1$ model (setting $a=1$)\footnote{Strictly speaking  
the $c=1$ limit is obtained by setting the deformation in (\ref{defmatr})  
to zero, which amounts to setting $f=1/(2\mu)$ in (\ref{bounce0aperturbative})  
and next re-expanding in $\mu\to\infty$. This will precisely   
reproduce the bounce factor of the $c=1$ matrix model.   
In particular, upon making this substitution, the {\it a priori}   
infinite series  
in (\ref{bounce0aperturbative}) will  
truncate to order $1/\mu^n$ for an integer value of the momentum $p=n$.}.  
In this case,   
the first line in the above expression coincides with the   
appropriate result quoted in \ref{bounceperturbative}. A less   
trivial consistency check can be obtained as follows. Setting   
$a=0$  brings us to the case discussed in \cite{igor}, where the   
coupling was identified as $M=(\mu\, f)^2-1/4$. The first   
interesting observation is that in the expansion of the bounce   
factor all odd powers of $\mu^{-1}$ are proportional to $a$ and   
therefore vanish in the limit $a\to 0$, in perfect agreement with   
\cite{igor}. Taking $a$ to zero in (\ref{bounce0aperturbative}) we   
obtain (including a term not written above)   
\bea   
\label{dkrbounce}  
R(p)&=&1+ \left(\frac{7}{24} p -\frac{1}{6}p^3 \right)M^{-1}   
\\  
&+& \frac{p(p-2)}{5760}\left(80p^4-128p^3+536p^2+128p+510)\right)M^{-2}  
+ \calo(M^{-3})\, . \nonumber   
\eea  
This reproduces the expansion of the bounce factor of Demeterfi,   
Klebanov and Rodrigues \cite{igor} (equations (12) and (13)).  
Notice that in the limit when the free fermion Fermi energy $\mu$ is   
vanishing, and one expands the bounce factor in $M$, the strength of the   
deformation in (\ref{defmatr}), the infinite series expansion   
in (\ref{dkrbounce}) truncates for even integer values of the   
momentum.  
  
\subsection{Correlators}  
  
As mentioned before, the building blocks of the   
partition function are correlators of the form $\langle {\calj}_p^n{\calj}_{-p}^n\rangle$. They can be computed directly with the methods  
developed by \cite{mpr}, but the calculations can soon become rather tedious.  
Instead, as in \cite{mooresg}, it is  technically simpler  
to compute   
correlators with an extra insertion of a zero-momentum vertex   
operator which we will denote by $A_n$. This section is dedicated   
to their evaluation after which we can proceed with the derivation of the   
partition function.  
  
\subsubsection{$\langle {\calj}_0 {\calj}_p^n {\calj}_{-p}^n\rangle$   
correlators}  
  
The $A_n$ amplitudes are related to the correlators we need for the   
partition function by the following equation:   
\be   
A_n=(\sqrt{\mu^2+M})^{-np}\frac{1}{\mu}\partial_{a}   
((\sqrt{\mu^2+M})^{np}{\cal R}_{n\to n})  
=(\sqrt{\mu^2+M})^{-np}  
\langle {\calj}_0 {\calj}_p^n {\calj}_{-p}^n\rangle\, . \label{an0}   
\ee  
where ${\cal R}_{n\to n}=\sqrt{\mu^2+M}^{-np}\langle {\calj}_p^n {\calj}_{-p}^n  
\rangle$ is the S-matrix element corresponding to the   
scattering of n tachyons of equal momenta into $n$ tachyons of equal momenta.  
  
This stems form the observation that, as in the case of the  $c=1$ model,   
in the bounce factor the dependence on the momentum $p$ arises in the   
combination  $p-i\mu a$, with the exception of the prefactor  
$\sqrt{\mu^2+M}^{-p}$. Hence we can trade again the differentiation with   
respect to $a$ for a differentiation with respect to the momentum.   
The latter, after partial integration, when acting on the Heaviside   
functions of the integrand,    
yields delta-functions and so, the net effect is to reduce the   
evaluation of (\ref{an0}) to the same simple algebraic computation   
according to (\ref{an}). Recall that in (\ref{an}) the insertion of   
the cosmological constant ${\calj}_0$ inside the correlator was   
done with the same means of differentiating with respect to $\mu$, and to the   
same end. Thus, (\ref{an0}) can be evaluated as in the $c=1$ case,   
using the definition of  (\ref{an}), where we insert the asymptotic   
expansion of the bounce factors (\ref{bounce0a}). At the expense of  
being too explicit but with the hope of exemplifying the  
simplification achieved by the combinatorial formula quoted in  
(\ref{an}) and obtained in  
\cite{mooresg}  we list the first few terms:  
\bea\label{cexp}  
-iA_1&=& R_0R^*_p-R_p R^*_0, \nonumber \\  
-iA_2&=& R_{2p}R^*_0-R_0R^*_{2p}-2R^2_pR^*{}^2_0+2R^2_0R^*{}^2_p,  
\nonumber  \\  
-i A_3&=&R_0R^*_{3p}-  
 R_{3p}R^*_{0}+3R_{p}R^*_{2p}-3R_{2p}R^*_p+9R_pR_{2p}R^*_0{}^2  
-9R^*_pR^*_{2p}R_0^2 \nonumber \\  
&+&9R_p^2R^*_pR^*_0-9R_pR^*_p{}^2R_0  
-12R_p^3R^*_0{}^3+12R_0^3R^*_p{}^3,  
\eea  
where $*$ represents complex conjugation.   
All we need to do at this point is to substitute the asymptotic  
expansion for the bounce factor (\ref{bounce0a}).   
We present only the first few genus zero correlators:  
\bea\label{correl}  
A_1&=& -\frac{a p^2}{\m(f^2+a^2)}, \nonumber \\  
A_2&=&-2!\frac{a p^4[(1-p)^2a^2+(2p-3)f^2]}{\mu^3(a^2+f^2)^3}, \nonumber \\  
A_3&=&-3!\frac{a p^6[(1-p)^3(3p-4)a^4+(13p^3-54p^2+78p-40)a^2 f^2  
+(3p-4)(3p-5)f^4]}{\mu^5(a^2+f^2)^5},\nonumber\\  
\dots\nonumber\\  
\eea   
The higher genera correlators correspond to subleading order terms in   
$1/\mu^{2n-1+h}$.  
  
We would like to comment on the main difference between the $c=1$  
correlators (which at genus zero are obtained by setting $f=0$ in   
(\ref{correl})), and the   
generic case with both $a, f$ non-vanishing. Namely, in the $c=1$ case  
all correlators  $A_n$ with $n\geq 2$ vanish for a special value of the   
momentum, $p=1$. The reason why this is happening is that for $p=1$, the   
$c=1$ bounce factor is simply $R(p=1)=1+\frac i{2\mu}$, and for integer  
momenta $R(p=n)$ is a degree $n$ polynomial in $1/\mu$.  
Substituting this  
into (\ref{cexp}) one finds that the highest power of $1/\mu$ for a   
given $n$ is $1/\mu^n$.  
However, according to KPZ scaling, these correlators should scale with   
$1/\mu^{2n -1}$. Thus, for $p=1$, all $A_n$ with $n\geq2$ must vanish.  
A short proof by induction shows that $p=1$ is a zero of order  
$n$ for the amplitude $A_n$.  
  
On the other hand, in the 0A case the bounce factor is an infinite series   
in $1/\mu$ (see (\ref{bounce0aperturbative})), and the previous argument does  
not apply anymore. Indeed, the correlators (\ref{correl}) have the   
right KPZ scaling, and are non-vanishing for $p=1$ as long as the RR flux   
is not turned off ($f\neq 0)$.  
It is also worth mentioning that even though there is a similar truncation  
of the bounce factor that takes place for the 0A bounce factor for even   
integer values of the momentum, this truncation happens only for $\mu=0$.  
In fact, the correlators $A_n$ are odd $2n+1$-point functions which   
vanish when $\mu$ (read $a$) is zero.

\subsubsection{The $\mu\to 0$ limit and $\langle {\calj}_p^n {\calj}_{-p}^n  
\rangle$ correlators}  
As it has been already discussed in the previous section, we obtain   
the building blocks of the partition function in the presence of   
momentum modes,  
$\langle {\calj}_p^n{\calj}_{-p}^n   
\rangle$, by performing the integration with respect to $a$ in   
(\ref{an0}).   
  
First, we notice that by setting the RR flux to zero ($f=0$) we   
reproduce the correlators of the $c=1$ model (\ref{tntn}), as expected.   
As shown in \cite{mooresg} and reviewed in section \ref{review},  
at zero RR flux the correlators acquire an expression that can be generalized for all $n$.   
  
For general values of the Liouville coupling $\mu$ and RR flux   
we have been unable to find a universal   
expression for all $n$ correlators. Interestingly, there is another   
limit\footnote{This  
limit was discussed recently by A. Kapustin \cite{kapustin}.} where such an   
universal expression can be   
found. The limit sends the cosmological constant to zero,   
$\mu \to 0$ (or equivalently $f\gg1$). In   
this limit, the correlators which enter in the genus zero   
0A partition function can be written as:  
\be \label{correlatorsn}  
\langle {\calj}_p^n{\calj}_{-p}^n \rangle=   
-n!(-1)^n q^{np-2n+2}(1-p)p^{2n}\frac{\Gamma(n(2-p)-2)}{\Gamma(n(1-p)+1)}\,   
,   
\ee  
where, after taking the limit $f\gg1$,  we reverted to the original notation   
$\mu f=q$, with $q$ the RR background flux of the two-dimensional type 0A string\footnote{Amusingly, the next-to-leading order term in $a/f$, or equivalently  
$\mu/q$, has also a universal expression:  
$$  
\langle {\calj}_p^n{\calj}_{-p}^n \rangle=   
-n!(-1)^n q^{np-2n+2}p^{2n}\bigg((1-p)\frac{\Gamma(n(2-p)-2)}  
{\Gamma(n(1-p)+1)}+\frac{\mu^2}{2q^2}\frac{\Gamma(n(2-p))}{\Gamma(n(1-p)+1)}
\bigg) \ldots$$ We were unable to organize the other subleading terms in a   
similar manner.  
}.   
We would like to stress that the limit $f\gg1$ should not be taken   
prematurely. Even though the correlators (\ref{correl})  
organize themselves in such a way that in the numerator, which is a  
polynomial in $f$, the highest and lowest order term in   
$f$ can be written as ratios of Euler $\Gamma$-functions while the  
rest of the terms have no apparent structure, as we perform the   
integral over $a$ all the terms in the numerator are equally contributing   
to the final result (\ref{correlatorsn}).

Our formula (\ref{correlatorsn}) reproduces known results in the literature.   
Namely, for the 2-point function we obtain:   
\be \langle {\calj}_p{\calj}_{-p} \rangle=\frac 12  
q^p p\, ,    
\ee  
which coincides with the results of  
\cite{jy,igor,dr}: see for instance eqn (15) in \cite{igor}.   
Recall that the correlators and $n$-point functions are related by   
multiplication   
with leg factors: $<\calj_p^n \calj_{-p}^n>={\cal R}_{n\to n}(p,\dots p; -p,\dots -p)  
q^{2np/2}$. Similarly, we find agreement for the 4-point function   
(eqn. (17) in \cite{igor})  
\be \label{4pt}  
\langle {\calj}_p^2{\calj}_{-p}^2 \rangle=q^{2p-2} p^4\, .   
\ee  
For comparison the $c=1$ 2-point function and   
4-point function, as given by eqns. (4.17) and (4.40) in \cite{moore},  
are:   
$\langle {\calj}_p {\calj}_{-p}  
\rangle=p\mu^p$, and  $\langle {\calj}_p^2{\calj}_{-p}^2  
\rangle=p^4 (p-1)\mu^{2p-2}$ respectively. The difference between the    
0A 4-point function \cite{jy,igor,dr}  
(\ref{4pt}) and the $c=1$ model result is reflected in the different  
dependence on $(1-p)$ encoded in the 0A generic formula (\ref{correlatorsn})  
vs. (\ref{c1cor}).  
  
Note that the role of the genus expansion which was originally played  
by $\mu$ is now played by $q$ in precise agreement with the KPZ scaling.   
An interesting point to address is that  
of the order of limits. In the  
original  works of \cite{igor,dr} the strategy was to set the  
cosmological constant to  zero at the beginning of the  
calculations. This was also suggested in works by Jevicki and Yoneya  
\cite{jy}. Here, and in the approximation considered by Kapustin,  
we have started with a nonzero cosmological constant (nonzero $a$) and  
obtained a formula in the limit of large flux  which is basically  
$f/a\gg 1$. In the end, we have found that the expression derived for the   
2n-point functions $\langle{\calj}_{p}^n{\calj}_{-p}^n\rangle$ is not  
sensitive to the order of limits. This independence of  
the order of limits hints to the existence of a deeper relation  
between the couplings $\mu$ and $q$ beyond the extreme limits when  
either of them is effectively zero.

\subsection{Partition function}  
The sum  
\bea \sum_n\frac{(\alpha^2 )^n}{2^n n!^2}\langle {\calj}_p^n{\calj}_{-p}^n \rangle   
\nonumber  
\eea  
can be performed after first differentiating twice with respect to   
$q$ and using (\ref{sumformula}). Thus, upon taking the limit $\mu\to 0$,   
the 0A string partition function admits an analytic expression  
\be  
\partial_q^2 Z=\partial_{q}^2 Z_{n=0}+  
(1-p)\sum_{n\geq 1}\frac{1}{n!}(-q^{p-2} p^2 \alpha^2)^n\frac{   
\Gamma(n(2-p))}{\Gamma(n(1-p)+1))}= -\ln\;q +(1-p)\ln(1-s)\label{part} \, ,  
\ee  
where, for us,   
\be \eqlabel{s} \frac{s}{(1-s)^{2-p}}=q^{p-2}p^2\alpha^2 \equiv z\,.   
\ee   
Let   
us contrast the current situation with the $c=1$ model   
\cite{mooresg,kkk}. While $z$ in the $c=1$ model could have been   
positive for $p>1$, or negative for $p<1$, in our case we see that   
$z$ is always positive. Moreover, now $z$ varies monotonically with   
$s$ for all $0<p<2$. Recall that in  the $c=1$ string one had to   
distinguish between a monotonic $z$ behavior with $s$ for $1<p<2$,   
and a non-monotonic one for $0<p<1$. In the latter case, $z$ was   
bounded by a critical value $Z_c$ reached for $(dz/ds)|_{Z_c}=0$ for   
$p<1$. In the vicinity of the extremum, one finds the susceptibility   
$\chi=\partial_\mu^2 Z$ being  
 proportional to $(z-Z_c)^2$, behavior that is characteristic to   
a $c=0$ system. The physical   
interpretation is that the $c=1$ field $X$ decouples by settling  
into the minima of the cosine potential corresponding to the turning on of   
the momentum modes. Thus $p=1$ is a critical   
point associated with the phase transition from the $c=1$ string   
to a $c=0$ model coupled to gravity. We will   
soon see that this decoupling is absent in our case.       
  
Returning to the two-dimensional 0A string, and similarly defining   
$\chi=\partial^2_q Z$, we find that $\chi$ obeys:  
\be\label{chi}  
q^{\frac{1}{1-p}}\, e^{\frac{1}{1-p}\, \chi} + \alpha^2 p^2  
q^{\frac{p(2-p)}{1-p}}\, e ^{\frac{2-p}{1-p}\,\chi}=1\,.  
\ee  
Sending $\alpha\to 0$ in the above expression brings us back to the expected  
answer  
\be  
\chi_{\alpha=0} = -\ln q\,.  
\ee   
Alternatively, we can directly explore the  limit  
$q\to \infty$ (instead of $\alpha\to 0$), by redefining   
$\chi=-\ln q+\hat \chi$, with $\hat\chi$ finite for large flux   
and constrained by  
\be\label{chi2}  
1=e^{\frac{\hat\chi}{1-p}}+\alpha^2p^2q^{p-2}e^{\frac{2-p}  
{1-p}\hat\chi}\,.  
\ee  
However, the limit that we are interested in is $\alpha\gg q$, or   
equivalently $q\to 0$. This regime can be probed by exploiting the analyticity   
of the equation (\ref{chi}) which allows us to re-expand the   
partition function around a background provided by the momentum modes.   
It is clear from (\ref{chi}) that sending $q\to 0$  
cannot be done without assuming that $\chi$ blows up at the same time. More  
precisely we need $\chi=-p\ln q +\tilde \chi$, with $\tilde\chi$ defined by  
\be\label{chi1}  
1=qe^{\frac{\tilde\chi}{1-p}}+\alpha^2 p^2 e^{\frac{2-p}{1-p}\tilde\chi}\, .  
\ee  
We can accomplish the re-expansion of the partition function   
in a regime where $\alpha\gg q$ by simply observing that the small expansion   
parameter $z$ in (\ref{part}) corresponds to $s\approx 0$, while  
a large $z$ corresponds to $s\approx 1 $. Therefore, to expand around   
large $z$, all that is needed is to  
replace the term $\ln(1-s)$ in (\ref{part}) by $\ln(s)\equiv\ln(1-t)$.   
Solving for $t$ yields  
\be   
\eqlabel{deft}  
t/(1-t)^{1/(2-p)}=z^{-1/(2-p)}\equiv y\, .  
\ee  
Using that $\tilde\chi(2-p)/(1-p)=\ln(1-t)$, the relation between the   
function $\tilde\chi$ and the new variable $y$ is given by  
\be\label{chiy}  
y=e^{-\frac 1{1-p}\tilde\chi(y)}-e^{\tilde\chi(y)}\,.  
\ee  
Furthermore, from  
\be  
F=\bigg(p^{\frac{2}{2-p}}\alpha^{\frac 2{2-p}}\bigg)^2  
\bigg[-\frac{py^2}2\ln(y\alpha^{\frac2{2-p}}p^{\frac2{2-p}})+y^2\frac{p-1}{2-p}  
\ln(\alpha p)+f(y)\bigg]\, ,  
\ee  
where $\partial_y^2 f =\tilde\chi(y)$,  
we finally arrive at the sought-after expression of the genus zero   
partition function of the two-dimensional type 0A string theory, in a momentum mode   
background:  
\bea  
 \!\!\!\!\!\!\!\!\!\!\!\!\!\!&F&\!\!\!\!=q^2(\frac{p}2\ln q +\frac{p-1}{2-p}\ln(\alpha p))\nonumber\\  
\!\!\!\!\!\!\!\!\!\!\!\!\!\!&-&\!\!\!\!\frac{\bigg(p^{\frac{2}{2-p}}\alpha^{\frac 2{2-p}}\bigg)^2}  
4\!\!\bigg[1\!\!+\!\!(-4p^2+4p\tilde\chi+4p-4)e^{\frac{p\tilde\chi}{p-1}}\!\!+\!\!(3p^2-3p-2p  
\tilde\chi)e^{\frac{2\tilde\chi}{p-1}}\!\!+\!\!(3p-2\tilde\chi p)e^{2\tilde\chi}\bigg]  
.\nonumber\\  
\eea  
{}From its definition (\ref{chiy}), one finds that $\tilde\chi \to 0$   
as $y\to 0$.  
Thus, for $q\to 0$, the partition function behaves as  
\be\label{fgz}  
F=\frac{(p-2)^2}{4}p^{\frac{2+p}{2-p}}\alpha^{\frac{4}{2-p}}+{\cal O}(\mu)\, .  
\ee  
The partition function of the sine-Liouville model at genus zero    
has the same KPZ scaling with $\alpha^{\frac{2}{2-p}}$, and   
$\alpha\gg1$. It appears then, that  
the 0A two-dimensional string theory at genus zero and  
in a background corresponding to the momentum modes condensation,    
becomes related to the sine-Liouville model, similarly  to the $c=1$ string.

\section{Phase diagram}\label{phase}  
In this section we consider the phase diagram in the $(\alpha, p)$ plane.   
A natural set of variables for addressing this question are  
$p$ and $z$. Basically, $z=0$ corresponds to the absence of   
momentum perturbation, that is, to $\alpha=0$. We are interested in   
the behavior of the partition function as  
$z\to \infty$ and in particular will look for singularities as we cover the range of couplings.   
  
Given that   
\be  
z=\a^2 \,p^2\, q^{p-2},   
\ee   
we are limited to the region of positive $z$ for all values of $p$.  
  
Varying $z$ from zero to infinity can be achieved by varying $s$.   
Note that the relation between $z$ and $s$ is monotonous. Indeed,   
using (\ref{s}) we conclude that   
\be  
\partial_s z= \frac{1+s(1-p)}{(1-s)^{3-p}}.  
\ee  
Monotonicity breaks when the above expression becomes zero, which   
happens for   
\be s_c=1/(p-1). \ee  
For $p<1$ we have $s_c<0$ and negative $s_c$ implies negative $z$   
through (\ref{s}) but this is outside the range of $z$, which we   
consider to be positive. For $1<p<2$, we have that $s_c>1$ which is   
also outside allowed domain for $s$ and $z$.

Thus, we verify that there is a monotonous relation between $z$ and $s$ and that it is possible   
to vary $z$ without   
obstruction in the full range $0\le z < \infty$ by taking $0\le s<1$. In the language of the coupling   
$\alpha$, this means that we can vary it in the range $0\le \alpha < \infty$ with no obstruction, as long as $0<p<2$. The expansion  
for small $\alpha$, that is, small $z$, is given by formula   
(\ref{part}), whereas the expansion for large $\alpha$ is given   
by,    
\be \eqlabel{largealpha} \chi=-p\ln q -\frac{1-p}{2-p}\ln(p^2\alpha^2)   
+\frac{1-p}{2-p}\sum_{n\geq 1}\frac   
1{n!}\frac{\Gamma(\frac{n}{2-p})} {\Gamma(\frac{n}{2-p}-n+1))}   
\bigg(-\frac{q}{p^{\frac 2{2-p}}\alpha^{\frac 2{2-p}} }\bigg)^n\, ,   
\ee   
where we have introduced the appropriate small parameter (\ref{deft}).   
In appendix \ref{app} we complement this analysis with a   
more explicit discussion.   
  
To conclude, let us present an alternative analysis of the phase   
structure of the partition function. Here we will follow some of the   
standard techniques for studying series convergence which where applied to the   
$c=1$ case in \cite{mooresg}. The main object is the function   
\be H(p;z)\equiv \sum\limits_{n=1}^\infty   
\frac{\G(n(2-p))}{n!\G(n(1-p)+1)}z^n. \ee  
The radius of convergence is   
\be \eqlabel{radius} |z|<R_c=\exp\left((p-2)\ln   
|p-2|-(p-1)\ln|p-1|\right).\ee  

There are basically four regions, recall that in our case   
$z=\a^2 \, q^{p-2}\,p^2\geq 0$ : \\  
I.) $0<p<2, \quad  0\le \a^2 q^{p-2}< R_c/p^2$\\  
II.) $2<p<\infty, \quad 0\le \a^2 q^{p-2}< R_c/p^2$\\  
III.) $0<p<2, \quad \a^2 q^{p-2}> R_c/p^2$\\  
IV.) $2<p<\infty, \quad \a^2 q^{p-2}> R_c/p^2$\\  
In contrast with the phase diagram of the $c=1$ model perturbed   
by momentum modes (Sine-Liouville), two of the   
phase space regions, distinguished by $0<p<1$ and $1<p<2$ have coalesced  
(recall that in our case $z$ stays always positive). As a   
consequence, the phase transition of the $c=1$ string in a momentum  
modes background to the $c=0$ model coupled to gravity,  
which took place at $p=1$, has disappeared from the phase diagram   
the 0A string.   
  
  
We have included regions II and IV for completeness. Region II has a   
singularity but it is expected since it corresponds to   
non-normalizable $\alpha$ perturbation, that is, an irrelevant   
perturbation which in the string theory diverges as $\phi\to \infty$   
rather than dying off. The partition function in region III is to be   
computed using eqn. (\ref{largealpha}). Remarkably similar formulas were   
obtained in \cite{mooresg} for regions II and IV.

\section{Conclusions}\label{conclusions}  
Let us comment on some aspects of our calculations  
and some interesting open problems.   
  
There are several approximations which we had to make in order   
to arrive at an analytic answer.   
One particular point that one would like to improve on is relaxing  
the condition of large flux. Note that in this sense we differ from previous results in the literature  
where the vanishing flux limit was taken \cite{yin,park}. We used perturbative techniques to arrive   
at an  expression for the genus zero two-dimensional 0A partition   
function perturbed by momentum modes, in the limit of vanishing   
cosmological constant $\mu$. Exploiting the analyticity of our result   
 we were able to probe regions characterized by arbitrary values of   
the RR flux $q$ and momentum modes coupling constant $\alpha$.   
We explicitly check the existence of a perturbative   
expansion around large values of $\alpha$, corresponding to a condensation  
of momentum modes. The phase diagram analysis shows that for momentum  
values below 2, such that the momentum mode vertex operator remains   
relevant, the phase transition to a $c=0$ system coupled to gravity  
is absent and there is no obstruction to turning on an arbitrarily  
large value of $\alpha$.   
It would be interesting to study the problem for generic values of $\mu, q$.  
One would hope that the analysis at  
intermediate values of $q/\mu$ would perhaps uncover a richer phase  
structure.    
  
We would like to point out the benefits of keeping the Fermi level  
$\mu$ non-vanishing in the intermediate stages of our calculation,   
even though ultimately we had to assume   
the limit $\mu\ll1$. Sending $\mu$ to zero prematurely would have   
left us with only one means of evaluating the two-dimensional 0A correlators   
$\langle {\calj}_{p}^n {\calj}_{-p}^n\rangle$, namely integrating the   
loop momentum following \cite{mpr}. Instead, keeping $\mu$ non-vanishing  
allows differentiating the correlators with respect to $\mu$, and   
subsequently turning a tedious integral into a simple algebraic  
expression, as in \cite{mooresg}.   
  
In a sense our calculation can be viewed as part of a more general conjecture   
mirroring that of Fateev, Zamolodchikov and Zamolodchikov   
\cite{fzz}. The FZZ conjecture states  (as presented in   
\cite{kkk}) that the $SL(2)/U(1)$ coset CFT is equivalent to the   
Sine-Liouville model, $c=1$ CFT coupled to a Liouville field, with   
the cosmological constant tuned to zero and the scale set by the   
winding mode of the $c=1$ field. It would be interesting to investigate the precise formulation   
of the conjecture in the presence of fluxes $q$.   
  
We hope that our results will shed light into the integrable   
structure of type 0A. In fact, we have partially studied the   
perturbation by momentum in the framework of the string equation and will report on our   
findings in an upcoming work \cite{0astringequation}.   
  
Recently \cite{ms} have discussed the finite temperature partition functions for   
0A and 0B establishing $T$ duality explicitly. It would be interesting to consider the   
extension of our work to the Euclidean case when the $X$ field lives in a circle as well as its 0B   
counterpart. We hope to return to some of the fascinating issues in perturbing two-dimensional   
string theories with momentum and winding operators.

\section*{Acknowledgments}  
We are grateful to O. Aharony, J. Davis, F. Larsen  and J. Walcher for discussions.   
We would like to thank J. Maldacena and N. Seiberg for illuminating comments and G. Moore   
and D. Kutasov for correspondence. LAPZ is grateful to Tel Aviv University and University of Iowa  
for hospitality during various stages of this project. This work is supported by DoE under grant  
DE-FG02-95ER40899.  
  
\appendix  
\section{The Lagrange Inversion Formula applied to the partition function}\label{app}  
In this appendix we showed that the main formula used in body of the   
paper repeatedly (\ref{sumformula} and \ref{part})  follows as a   
direct application of a theorem due to Lagrange \cite{lagrange}. Our   
discussion follows \cite{ww,generating}.   
  
{\it Theorem.} (The Lagrange Inversion Formula)  Let $f(z)$ and $\phi(z)$ be functions of $z$ analytic   
on and inside a contour $\calc$ surrounding a point $a$, and let $t$   
be such that the inequality   
\be |t\phi(z)|<|z-\a|, \ee   
is satisfied at all points $z$ on the perimeter of $\calc$. Then the   
equation  
\be \xi =\alpha+t\phi(\xi), \ee  
as an equation in $\xi$ has one root in the interior of $\calc$; and   
further any function of $\xi$ analytic on and inside $\calc$ can be   
expanded as a power series in $t$ by the formula  
\be \eqlabel{lagrange} f(\xi)=f(\a)+\sum\limits_{n=1}^\infty   
\frac{t^n}{n!}\,\,\,\frac{d^{n-1}}{d   
x^{n-1}}\bigg[f'(x)(\phi(x))^n\bigg]_{x=\a}. \ee  
The case we are interested is basically   
\be y=1-z\, y^a, \ee  
In the formula \ref{lagrange} we simply have $f(y)=\ln(y)$ and   
$\phi(y)=y^a$ and obtain  
\be \ln y = -z + \frac{2a-1}{2}z^2 -\frac{(3a-1)(3a-2)}{6} z^3 +   
\frac{(4a-1)(4a-2)(4a-3)}{24} t^4 \ldots \ee  
which can be rewritten as    
\be \eqlabel{formula} \ln y =\sum\limits_{n=1}^\infty   
\frac{\Gamma(na)}{n!\Gamma(n(a-1)+1)}(-z)^n, \qquad {\rm with }   
\quad y=1-zy^a. \ee  
This is the formula used in the main body of the paper   
(\ref{sumformula}) and (\ref{part})  with the minor substitution of   
$y=1-s$ and for the case of $a=2-p$.   
  
Let us now discuss the regime of validity of the above expression   
and its possible continuation.  The above expansion \ref{formula} is   
valid for   
\be |z|<|(a-1)^{a-1}a^{-a}|,\ee  
which coincides with the radius of convergence given in section   
\ref{phase} by equaiton \ref{radius}. Having identified the series   
in $z$ with $\ln y$, one has a perfect analytic expression near   
$y=1$ for the partition function. Now we can analytically continue   
the natural logarithm. The only problem is with the branch cut   
$(-\infty,0]$. However, as explained in the main body, we are   
interested in $z\in [0,\infty)$  which corresponds to $y\in (0,1]$.   
Recall that the singularity in the $c=1$ case reviewed in section   
\ref{review} appears because $z$ takes negative values for $p<1$.


\end{document}